\documentclass{article}

\usepackage{arxiv}

\usepackage[utf8]{inputenc} 
\usepackage[T1]{fontenc}    
\usepackage{hyperref}       
\usepackage{url}            
\usepackage{booktabs}       
\usepackage{amsfonts}       
\usepackage{nicefrac}       
\usepackage{microtype}      
\usepackage{lipsum}		
\usepackage{graphicx}
\usepackage{natbib}
\usepackage{doi}

\title{Analysis of Heuristic and Digital Filters as Applied to Video-oculography Signals}

\author{ {\hspace{1mm}Mehedi H. Raju}\thanks{corresponding author} \\
	Texas State University\\
    601 University Drive\\
    San Marcos, Texas, 78640, USA\\
	\texttt{m.raju@txstate.edu} \\
	\And
	{\hspace{1mm}Lee Friedman} \\
	Texas State University\\
    601 University Drive\\
    San Marcos, Texas, 78640, USA\\
	\texttt{lfriedman10@gmail.com} \\
	\And
	{\hspace{1mm}Troy M. Bouman} \\
	Department of Mechanical Engineering-\\Engineering Mechanics\\
    Michigan Technological University\\
    1400 Townsend Dr.\\
    Houghton, MI 49931, USA \\
	\texttt{tmbouman@mtu.edu} \\
	\And
	{\hspace{1mm}Oleg V. Komogortsev} \\
    Texas State University\\
    601 University Drive\\
    San Marcos, Texas, 78640, USA\\
	\texttt{ok11@txstate.edu} \\
}

\renewcommand{\shorttitle}{Analysis of Heuristic and Digital Filters as Applied to VOG}

\begin{document}
\maketitle

\begin{abstract}
In 1993, \cite{stampe} suggested two ``heurisitic'' filters that were designed for video-oculography data. Several manufacturers (e.g., SR-Research, Tobii T60 XL and SMI)  have employed these filters as an option for recording eye-movements.  For the EyeLink family of eye-trackers, these two filters are referred to as standard (STD) or EXTRA.  We have implemented these filters as software functions.  For those who use their eye-trackers for data-collection only, this will allow users to collect unfiltered data and simultaneously have access to unfiltered, STD filtered and EXTRA filtered data for the exact same recording.  Based on the literature, which has employed various eye-tracking technologies, and our analysis of our EyeLink-1000 data, we conclude that the highest signal frequency content needed for most eye-tracking studies (i.e., saccades, microsaccades and smooth pursuit) is around 100 Hz, excluding fixation microtremor. For those who collect their data at 1000 Hz or higher, we test two zero-phase low-pass digital filters, one with a cutoff of 50 Hz and one with a cutoff of 100 Hz.  We perform a Fourier (FFT) analysis to examine the frequency content for unfiltered data, STD data, EXTRA filtered data, and data filtered by low-pass digital filters.  We also examine the frequency response of these filters.  The digital filter with the 100 Hz cutoff dramatically outperforms both heuristic filters because the heuristic filters leave noise above 100 Hz.    In the paper we provide additional conclusions and suggest the use of digital filters in scenarios where offline data processing is an option.
\end{abstract}


\section{Introduction}
\label{intro}

In 1993, \cite{stampe} suggested two ``heurisitic'' filters that were designed for video-oculography data. Several manufacturers\footnote{SR-Research (Ottawa, Canada), Tobii (Stockholm, Sweden) and the XVIEW system from SMI (out of business). For Tobii, we are referring to the XL 60 specifically.}  have employed these heuristic filters as an option for recording eye-movements.  Going forward, our remarks will be based on data from our own EyeLink 1000, from which we have data and hands on experience.   The EyeLink eye-tracking devices (EyeLink 1000 and 1000 plus, and EyeLink II) in particular, are a very popular family of eye-trackers used for research that have been cited in over 10,000 publications\footnote{see \url{https://www.sr-research.com/full-publications-list}}.  This journal has published 87 papers using an EyeLink system.  Nonetheless, we recognise that other manufacturers use may these filters and our findings may apply to them.

In the EyeLink family of eye-trackers, these two filter options are referred to as standard (STD) or EXTRA. EXTRA takes as input STD filtered data. (In other cases, the filters are referred to as level 1 and level 2 filters.)  We have implemented these filters as software functions.  For those who use their eye-trackers for data-collection only, this will allow users to collect unfiltered data and simultaneously have access to unfiltered, STD filtered and EXTRA filtered data for the exact same recording.  Until now, the only way to get unfiltered data from an EyeLink device is to collect it unfiltered. If, instead, the data are collected using the STD or EXTRA filters, access to unfiltered data is lost.  As far as we are aware, no one prior to us has created software functions which perform these filtering steps.  With our software functions\footnote{Written by Dr. Friedman and available in MATLAB (Natick, Massachusetts) at  \url{https://digital.library.txstate.edu/handle/10877/16136}. Information about how the filters were tested is also included.  Python versions are available as part of our Supporting\_Material.zip.  See below.}, users can now collect their data unfiltered and simply apply our filter functions to the unfiltered signal to get either an STD or EXTRA filtered signal.  Such an approach allows one to judge which filter level is best while retaining access to all other filtered and unfiltered data.  We can also imagine a scenario in which some analyses are performed on the unfiltered data and some analyses can be performed on filtered data.  However, as we will show, if the goal of a filter is to retain the wanted signal and remove the unwanted signal, these heuristic filters dramatically under-perform digital low-pass filters, and we recommend against the use of the heuristic filters going forward.  In 1993, \cite{stampe} rejected ``linear'' filters:

\begin{quote}
\quad ``Eye-tracking systems with analog outputs such as EOG (electrooculography) or scleral/limbus reflection devices permit high sampling rates and resolution, although their noise levels are fairly high. For these systems, linear filters are often used to remove noise (Inchingolo \& Spanio, 1985). Because of the low sampling rate of pupil-tracking systems, linear filters would smooth position data excessively, making saccade detection difficult or impossible.\\

\quad Instead of linear filtering, template-matching logical filters may be utilized. These filters compare each data sample with neighboring samples and modify or pass the sample accordingly. They function well at low sample rates and add little or no delay to the data processing. Their template-matching characteristics make them ideal for removing impulse noise and detecting saccades or fixations in the data.'' (Page 138, left column).
\end{quote}

This statement may have been true when \cite{stampe} made it in 1993.  EyeLink devices at that time had sampling rates of 250 Hz.  But as of today, this statement no longer holds.  As is evident in Table \ref{tab:Hz}, for the search coil studies, the lowest sampling rate was 200 Hz and the highest was 1000 Hz. The median sampling rate for the search coil studies was 1000 Hz.  For EOG, the lowest rate was 100 Hz and the highest rate was 10,000 Hz.  The median sampling rate for EOG studies was 762 Hz.  Only one research group employed a 10,000 Hz sampling rate for EOG \citep{Hz_8,Hz_9}.  All of the other papers reported sampling rates of 1024 or less.  Looking at the sampling rates in Table \ref{tab:Hz}, and considering that the EyeLink systems now use sampling rates of 500 Hz, 1000 Hz or 2000 Hz, there is no basis for saying that the EyeLink sample rates are low at the current time.

\begin{table*}[htbp]
    \caption{Sampling Rates for Scleral Search Coil (SSC) and EOG Methods}
    \centering
    \begin{tabular}{l|l|l|l}
    \hline
        Articles & Year & Method & Sampling Rate (Hz) \\ \hline
        \citep{Hz_1} & 1998 & SSC & 500 \\ \hline
        \citep{Hz_1A} & 2001 & SSC & 200 \\ \hline
        \citep{Hz_2} & 2002 & SSC & 500 \\ \hline
        \citep{Hz_3} & 2007 & SSC & 1000 \\ \hline
        \citep{Hz_4} & 2009 & SSC & 1000 \\ \hline
        \citep{Hz_5} & 2013 & SSC & 1000 \\ \hline
        \citep{Hz_6} & 2013 & SSC & 1000 \\ \hline
        \citep{Hz_7} & 1996 & EOG & 200 \\ \hline
        \citep{Hz_8} & 2005 & EOG & 10,000 \\ \hline
        \citep{Hz_9} & 2005 & EOG & 10,000 \\ \hline
        \citep{Hz_10} & 2019 & EOG & 500 \\ \hline
        \citep{Hz_11} & 2020 & EOG & 1024 \\ \hline
        \citep{Hz_12} & 2021 & EOG & 100 \\ \hline
    \end{tabular}
    \label{tab:Hz}
\end{table*}

We think that when Stampe \citep{stampe} says ``linear filter" he means some kind of digital low pass filter.  We note that \cite{stampe} never assessed the frequency-response of these heuristic filters.  All filters have a frequency-response, even heuristic filters.  We will demonstrate that digital low pass filters dramatically outperform the heuristic filters for unfiltered EyeLink 1000 signals.  And we judge it very unlikely that these filters would do anything but improve saccade detection.

Before we explain our approach we want to review the literature on what frequencies are required to preserve the events of interest (saccades and fixation tremor).

\subsection{Review of what is know about the frequency content of Saccades}

We present our analysis of the literature in Table \ref{tab:saccade}. One potentially relevant paper was not included in our table \citep{samplingfrequencyeffect}.  The signals (EOG and photoelectric) were analog signals.  These analog signals were filtered first with low-pass analog filter at 30 Hz. Subsequently, the signals were digitally filtered with a low-pass filter with a cutoff of 70 Hz.  This creates a very complex situation, and we don't think that statements about frequencies required to preserve saccade peak velocity are useful given the insertion of this analog filter and that is why the paper was not included.

We also exclude \citep{atleast200hz}.  Their paper was based on EOG signals which were analog filtered with a cutoff at 100 Hz.  Any further statements about the effects of other digital low pass filtering are confounded by the presence of the analog filter.

From Table \ref{tab:saccade}, despite the difference in recording and other methods, it would appear that the literature  supports the notion that frequencies from 0-125 Hz are sufficient to preserve saccade characteristics.

\begin{table*}[!ht]
\caption{Frequency content of saccades}
\label{tab:saccade}
\begin{tabular}{p{0.2\linewidth} | p{0.2\linewidth} |p{0.5\linewidth}}
\hline
Article      & Methods &  Findings\\
\hline
\citep{bahill80Hz}  & Photoelectric techniques &  For noisy data, a bandwidth of 0-125 Hz was required to record saccades.  Also, a sampling rate of 1000 Hz was suggested.\\   \hline
\citep{250Hz} & Video-based infrared eye-tracker & Sampling rate should be 250 Hz. With a sampling rate of 250 Hz one can only resolve signals in the 0-125 Hz range. Therefore, frequencies up to 125 Hz preserved saccades.  They did not investigate low sampling frequencies. \\ \hline
\citep{50Hz}  & VOG and Search coil & Saccadic eye movements of $>=5^o$ amplitude are bandwidth limited up to a frequency of 25 to 30 Hz. \\ \hline
\citep{mack} & Synthetic saccades & Signals sampled as low as 240 Hz allow for good reconstruction of peak velocity.  With 240 Hz, the frequencies that can be evaluated are 0-120 Hz.\\\hline
\end{tabular}
\end{table*}

\subsection{Review of what is known about the frequency content of fixation tremor}

Video-based methods are not typically used to assess tremor and probably should not be:

\begin{quote}
``Due to OMT’s [``Ocular microtremor"] small amplitude and high frequency, the most accurate and stringent way to record it is the piezoelectric transduction method.
\ldots
Whereas the piezoelectric sensor can measure both microsaccades and OMT, video systems such as EyeLink II (SR Research) can measure microsaccades accurately, but do not have the resolution necessary to measure OMT.'' \citep{mccamy}
Information in brackets was added to the quote.
\end{quote} 

Nonetheless, we reviewed the literature on the frequency content of microtremor in Table \ref{tab:tremor} for future reference. 
None of the papers listed used video-oculography (VOG) for microtremor recording.  In Table \ref{tab:tremor}, the median low frequency bound for tremor is 50 Hz, and the median high frequency bound is 95 Hz. 

\begin{table*}[!ht]
\caption{Frequency Ranges of Fixation tremor}
\label{tab:tremor}
\begin{tabular}{p{0.2\linewidth} | p{0.2\linewidth} | p{0.5\linewidth}}
\hline
Article & Methods & Findings\\
\hline
\citep{adler1934influence}  & Mirror on EyeBall &  Tremor has frequency components greater than 40 Hz \\ \hline
\citep{ratcliff50}  & Mirror attached on contact lens &  Tremor has frequency ranges 30-70 Hz \\ \hline
\citep{ditchburn1953involuntary}  & Contact lens &  Tremor has frequency ranges 30-80 Hz \\ \hline
\citep{riggs1954motions}  & Mirror attached on contact lens &  Tremor has frequency ranges 60-80 Hz \\ \hline
\citep{matin1964}  & Mirror attached on contact lens using infrared light &  Tremor has frequency of 85 Hz \\ \hline
\citep{bengi1968fixation}  & Accelerometer fixed to a contact lens &  Tremor has frequency components greater than 40 Hz \\ \hline
\citep{shakhnovich1973omt,davies1978transducer}  & Piezoelectric strain gauge &  Tremor has frequency of 100 Hz \\ \hline
\citep{bengi1973studies} & Piezoelectric strain gauge & Tremor's frequency could extend to 150 Hz \\ \hline
\citep{MiniatureEM73} & Search coil (monkey)  & Tremor has frequency ranges 50-100 Hz \\ \hline
\citep{abakumova1975investigation}  & Piezoelectric strain gauge &  Tremor has frequency of 96 Hz \\ \hline
\citep{coakley1977ocular}  & Piezoelectric strain gauge &  Tremor has frequency of 101 Hz \\ \hline
\citep{michalik1983okulare}  & Piezoelectric strain gauge &  Tremor has frequency of 87.4 Hz \\ \hline
\citep{ezenman1985power}  & Corneal Reflection & Tremor has frequency components greater than 40 Hz \\ \hline
\citep{bolger1999dominant} & Piezoelectric strain gauge & Tremor's average frequency is 83.68 Hz (ranged from 70-103 Hz)\\ \hline
\citep{spauschus1999origin} & 2 accelerometers glued to the eyes & Spectral peaks of micro-tremor at observed a low (to 25 Hz) and high (60-90 Hz)\\ \hline
\citep{HKD} & DPI (human); Search coil (Monkey) & Tremor's average frequency is approximately 80~Hz in the range (50-100 Hz)\\ \hline
\citep{bowers2019effects} & Retinal Movies & Tremor has frequency ranges 50-100 Hz \\ \hline
\end{tabular}
\end{table*}

In the present study we collected unfiltered data with an EyeLink 1000 eye-tracker at a sampling rate of 1000 Hz. Eye movements were recorded during tasks involving fixation, random saccades, large horizontal saccades and smooth pursuit.  We implement the ``heuristic" filters used in our EyeLink 1000 \citep{stampe} as simple software functions.  We assess signal frequency content for fixations, saccades, and catch-up saccades that occur during smooth pursuit. We analyze the impact of both heuristic and digital filters on microsaccades, catch-up saccades, and saccades with post saccadic oscillation (PSO). We study the effects of filters on the saccade peak velocity as well.  We present and compare the frequency response of the heuristic and low-pass digital filters.  We offer our recommendations as to how users of EyeLink devices should proceed.

\section{Methods}
\label{method}

\subsection{Subjects}
We recorded a total of 23 unique subjects (M=17/ F=6, median age = 28, range = 20 to 69 years). From the total number of unique participants, 14 had a normal (not-corrected) vision, and 9 had a corrected vision (7 glasses, 2 contact lens). They were recruited from laboratory personnel, undergraduates taking a class on computer programming, and friends of the experimenters. The Texas State University institutional review board approved the study, and participants provided informed consent.

We report on two datasets, the first dataset, ``Fix-Hor'', contains data when subjects were either fixating on a central target (Fix) or making large horizontal saccades (Hor).  The second dataset, ``RS-SP'' contained data when subjects viewed random saccade targets or performed smooth pursuit.  The Fix-Hor dataset consisted of 15 subjects. The RS-SP dataset consisted of 9 subjects.  The 9 subjects in the RS-SP dataset were a subset of the 15 subjects in the Fix-Hor dataset.

\subsection{Eye Movement Data Collection}
Eye movements were collected with a tower mou-nted EyeLink 1000 eye tracker (SR Research, Ottawa, Ontario, Canada). 
The eye tracker operated in monocular mode capturing the participants dominant eye (9 left eye, 14 right eye). During the collection of eye movements data, each participant's head was positioned at a distance of 550 millimeters from a 19'' (48.26cm) computer screen (474x297 millimeters, resolution 1680x1050 pixels), where the visual stimulus was presented.  The sampling rate was 1000 Hz.

For the fixation task, subjects were presented with a single fixation point (white circle, $0.93^o$) as the visual stimulus. The point was positioned in the horizontal middle of the screen and a vertical angle of  3.5$^o$ above primary position. Participants were instructed to fixate on the stationary point stimulus for a period of 30 seconds. 

For the horizontal saccade task, there were 2 targets, one at $-15^o$ and one at $+15^o$. As soon as one target became visible, the subject was instructed to look at the target.  They fixated at that point for 1.0 sec.  After 1 sec, the old target disappeared and the new target, on the other side of the display, appeared.  This went on until both targets were viewed 50 times.  For the analysis of this data, we only analyzed saccades that were $>22^o$.

During the random saccade task, subjects were instructed to follow the same target on a dark screen as the target was displaced at random locations across the display monitor, ranging from ± 15$^o$ and ± 9$^o$ of visual angle in the horizontal and vertical directions, respectively. The random saccade task was 30 seconds long. The target positions were randomized for each recording.
The minimum amplitude between adjacent target displacements was 2$^o$ of visual angle. The distribution of target locations was chosen to ensure uniform coverage across the display. The delay between target jumps varied between 1 sec and 1.5 sec (chosen randomly from a uniform distribution).

During the smooth pursuit task, subjects were instructed to follow a target on the dark screen as the target moved horizontally from center to right. This ramp was followed by a fixation (length between 1 and 1.5 sec). This was followed by another ramp from the right to the left of the screen, then another fixation, etc.  The rest of the task was a series of left-to-right and right-to-left ramps with fixations interposed.  The target was moving at velocities  of either $5^o$/sec,  $10^o$/sec, or  $20^o$/sec. For each speed there were 5 continuous leftward and 5 rightward ramps per set.  The order of the velocity sets was random for each participant. There was a 15 sec fixation period at the beginning of the task and between each set. The whole recording was 120 seconds long.

\subsection{Signal Processing}
\subsubsection{Signal Processing of Fixation Data} \label{segemntselection}
All fixation recordings lasted 30 seconds (30,000 samples).  We used these fixation periods to create amplitude spectra and to determine the frequency response of the several filters discussed below.  The segment selection was a two step process. First, from each recording we chose non-overlapping segments of 2048-samples from the unfiltered recording for each subject. We calculated the velocity with a six point difference approach using $t_{+3}$ and $t_{-3}$ \citep{bahilltwopoint}. We rejected any segment which contained velocities above 25 deg/sec to reduce the possibility of saccades or other fast events in our segments. For four of the 16 subjects, we could not find a single segment of 2048 samples that met our criteria. For the remaining subjects we found from 1 to 4 segments.  We  use these 2048-sample segments for our Fourier analysis. Using the fast Fourier transform (FFT), the ratio of sampling rate to the segment size (i.e., block size) determines the frequency resolution.  For 2048-sample segments we could discriminate 1024 different frequencies from 0 to 500 Hz.  Since these analyses were quite noisy, we decided to separate the 2048-sample segments into 8 256-sample segments.  This would still give us reasonable frequency precision of approximately 4 Hz, and it would produce more segments to average (we had 27 2048-sample segments and these produced 216 segments of 256 samples across which to average).  Note that we did our averaging using the magnitude and phase-shift spectra rather than averaging the complex FFT data.

\subsubsection{Selection of Saccade, Catch-up Saccade and Microsaccade Exemplars}
We wanted to have multiple exemplars of saccades with PSOs, catch-up saccades (CUS) and microsaccades.  For the saccade examples, we used the random saccade task. For the CUS, we used the smooth pursuit dataset.  Microsaccades were selected from the fixation dataset.  Only one of each exemplar is presented in this paper.  Other examples are available (see footnote 4 below).

\subsection{Filter Functions implementing the STD and the EXTRA EyeLink 1000 filters}
EyeLink 1000 eye-trackers have three filter settings: OFF, Standard (STD), and EXTRA \\ \citep{stampe}. For OFF, no filtering was performed. The STD filter removes one-sample noise spikes.  The EXTRA filter removes two-sample noise spikes. Dr. Friedman used the \cite{stampe} paper to develop these filter functions in MATLAB\footnote{These functions, and additional information about their testing, is available at \url{https://digital.library.txstate.edu/handle/10877/16136}.  Download FilterFunctionsETC.zip. For python versions, go to the same website and download the Supporting\_Materials.zip}.

\subsection{Digital low pass filter design}
\label{ZPLP}
Below, when we refer to a ``cutoff'' frequency, we are referring to the -3db point, which is standard in the signal processing literature.  At the -3db point, signals are reduced by 50\%.  

To remove noise from our eye tracking signal we chose Butterworth (BW) low-pass filters for all digital filtering.  The BW filter is maximally flat in the pass-band and has no ripples in the stop-band. Also BW filters do not have any linear phase response in contrast to finite impulse response filters \citep{mack}. Mack et al go on to say:

\begin{quote}
    ``A general observation from the best filter list is the increasing prevalence of BW filters at higher sampling rates, ending in a total absence of other filter types at 1 kHz. This can be explained by considering the smoothness of the signal. At higher sampling rates more noise is present in the data. Such high frequency noise can be more efficiently suppressed by the steeper roll-off of the BW filters compared to \ldots''\citep{mack}, page 2159.
\end{quote}

Typically, low-pass digital filters can have phase effects.  A zero-phase filter can be constructed by first passing the signal through the filter in the forward direction, then the function reverses the filtered sequence and runs it back through the filter. Zero-phase filters will have a somewhat different frequency response compared to the component filter.  We estimate the frequency response of these zero-phase filters in our analysis.  Based on viewing initial results, we created two zero-phase low pass (Z-LP) filters, one with a cutoff at 100 Hz (Z-LP100) and one with a cutoff at 50Hz (Z-LP50). The usefulness of these two filters is analyzed below.  Ultimately, we recommend the use of the Z-LP100 filter and not the Z-LP50 filter.

\subsection{Fourier Analysis of Fixation}\label{fourieranlysis}
We began our analysis with sets of fixations that are (1) unfiltered, (2) filtered using the STD filter function,
(3) filtered using the EXTRA filter function, (4) filtered with a zero-phase Z-LP100 digital low pass filter and (5) filtered with a Z-LP50 low pass filter.

We were interested in comparing the amplitude spectra (and phase shift spectra) across these five signals.  For this we use Fourier (Fast Fourier Transform, FFT \citep{FFT}) methods to provide amplitude spectra (and phase shift spectra). We have 216 fixation segments that are 256 samples in length.  

We start by detrending each segment with a 2$^{nd}$ order polynomial. The residuals of these polynomials have mean zero. These detrended signals are then windowed with a Hann window. We then perform an FFT of these detrended and windowed data. The spectra have 128 frequency components, ranging from  0 to 500, with increments of 500\/128 = 3.90625 Hz.

All amplitude (and phase shift) spectra displayed in this report are averages across all 216 256-sample-length signals.

\subsection{Signal Frequency Content Analysis}
\begin{figure}[htbp]
\centering
\includegraphics[width=0.7\textwidth]{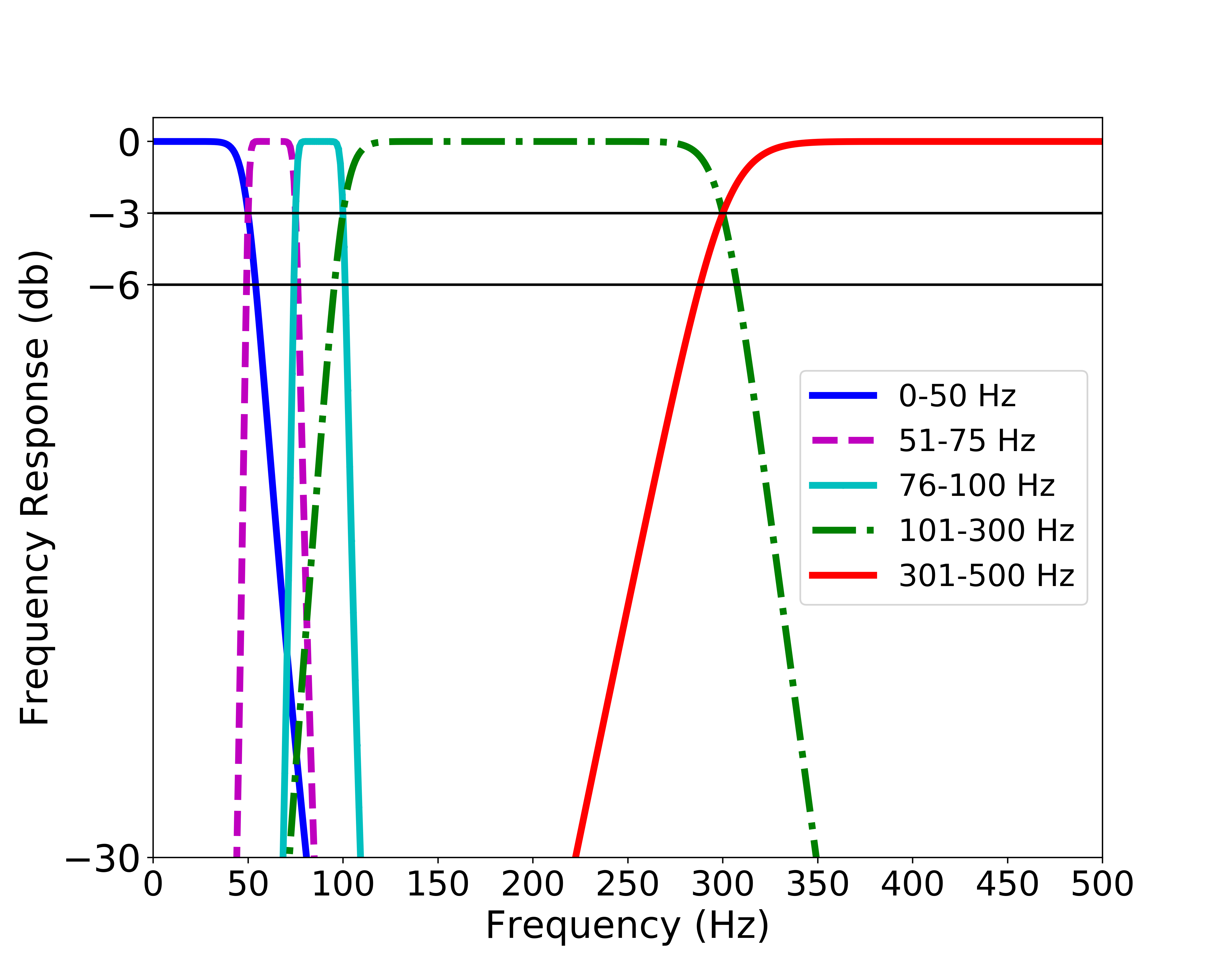}
\caption{Frequency response of different frequency bandwidths}
\label{fig:1}
\end{figure}

We wanted to assess the bandwidth of frequencies for saccades, CUS and microsaccades.  We created bandwidths (see Fig. \ref{fig:1}, 0-50 Hz, 51-75 Hz, 76-100 Hz and 101-300 Hz, 301-500 Hz).  These were created using very sharp high-pass, low-pass and band pass Butterworth-style filters (always zero-phase and order = 7).  The 50 Hz cutoff was implemented based on the findings of \citep{50Hz} that suggest that saccades generally have frequencies in the 0-50 Hz range.  Although there was some leakage in each filter, it was clear that our procedure does create sharp distinctions between bands.

More examples are available as part of our supplementary material.

\subsection{Estimation of Filter Frequency Response}
We wanted to estimate the frequency response of each of our four filters.  For each filter, we start with an FFT analysis of unfiltered data.  Let us label this FFT as ``A''. We also had FFT spectra for each filtered dataset. Let us label the FFT of the filtered signal as ``B''.  The frequency response for the filter was found by computing the ratio of $B/A$.

\subsection{Study of the Effects of Filtering on Saccade Peak Velocity} \label{PkVel}
Several prior reports have discussed potential effects of various filters on the measured peak velocity of saccades \citep{samplingfrequencyeffect,LPFeffect}.  To study this, we identified saccades in all the random saccade data from the RS-SP dataset.  The saccades were detected with an updated version of our previously published event detection software \citep{MNH}.  There were 9 subjects and 5 filter conditions per subject.  The median number of saccades per study was 209 (178 to 253).  Every saccade detected was plotted on a chart with the ordinate as log(peak velocity) and the abscissa as log(saccade amplitude).  We noted 2 clusters of points in the charts, one consisted of saccades less than or equal to $4^o$ (as logarithm, 1.39) and one with larger amplitude saccades.  The peak velocity in both clusters increased linearly with amplitude, but the slope of the fitted line was substantially higher in the smaller saccade group.  For each recording, we used a robust fit regression algorithm to measure the slope of both clusters.

For each cluster, the effect of filter type was assessed with a mixed model ANOVA approach with filter type as a fixed effect and subject as a random effect. A statistically significant filter effect was examined for pairwise comparisons using the Bonferroni method to control for multiple comparisons.

We also had a set of very large saccades from our horizontal saccade task where subjects were required to alternately fixate on two targets $30^o$ apart.  From this task, we chose saccades $>22^o$. We analyzed saccade amplitude and peak velocity using a mixed model ANOVA with filter type as a fixed effect and subjects as a random effect.  Once again, if we found a statistically significant filter effect, we would examine pairwise comparisons using the Bonferroni method to control for multiple comparisons.

\section{Results}
\label{result}

\subsection{Examining the effects of the various filters on a fixation time-series}
\label{FiltEffects}

\begin{figure*}[!ht]
\centering
\includegraphics[width=0.85\textwidth]{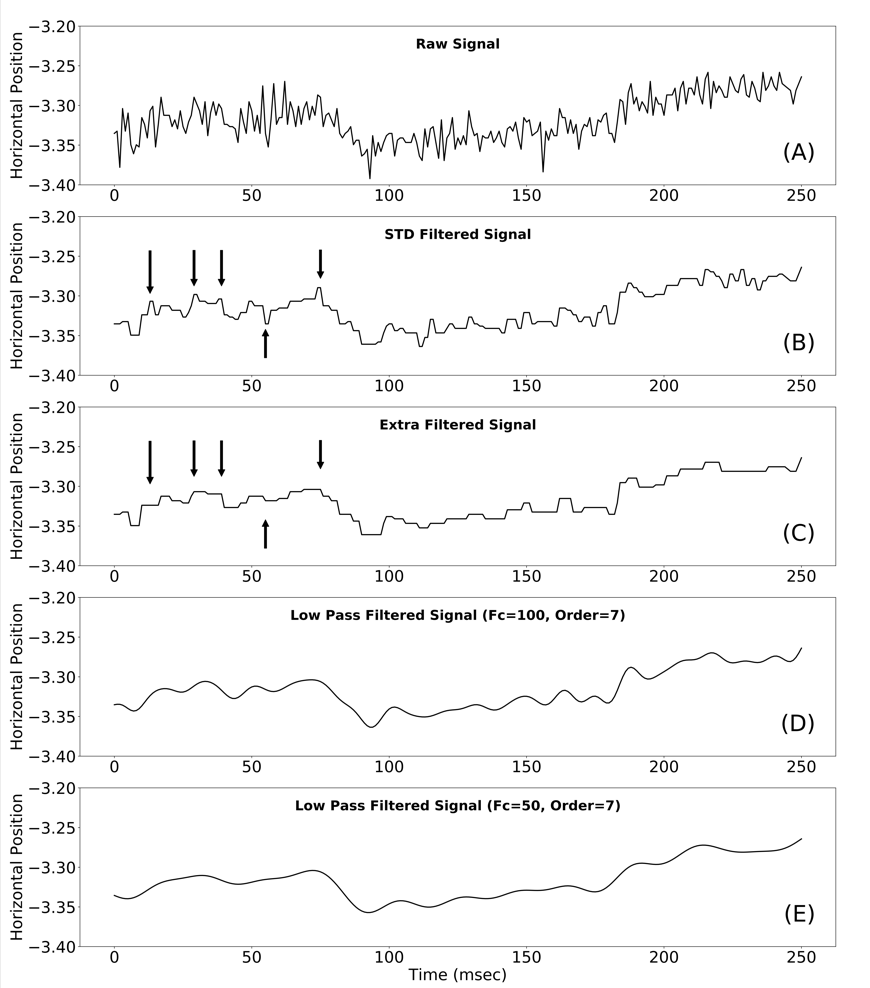}
\caption{Comparison of an unfiltered signal with filtered versions of the same signal.
(A) A fixation that is 250~ms in duration and is unfiltered.
(B) Signal in (A) has been filtered using our STD filter function. Arrows point to ``two-sample noise spikes" that the next filter (EXTRA) was designed to eliminate.
(C) Signal in (A) has been filtered using EXTRA filter function (which takes STD filtered data as input).  This filter was designed to remove ``two-sample noise spikes" indicated with arrows in (B). Note the absence of such spikes in (C).
(D) We filter the unfiltered signal with our Z-LP100 filter.
(E) We filter the unfiltered signal with our Z-LP50 filter.}
\label{fig:2}
\end{figure*}

In Fig. \ref{fig:2}, we present a single unfiltered signal segment (plot (A)).  In plot (B) we present the same signal as in plot (A) after it has been processed by our STD filter function.  One can see that the signal in plot (B) is much less noisy than that in plot (A). Many one-sample spikes have been removed. The arrows indicate the position of two-sample spikes. In plot (C) we present the same signal as in plot (B) after it has been processed by our EXTRA filter function.  Note the absence of the two-sample spikes in this signal.  In plot (D), we present the results of filtering the unfiltered signal with our Z-LP100 filter.  This signal has less high frequency noise than the STD and EXTRA filtered signals.  In plot (E) we present the results of filtering the unfiltered signal with our Z-LP50 filter. Note that it appears to have even less noise than the Z-LP100 filter, as would be expected.

\subsection{Fourier analysis of the unfiltered and filtered signals}
\label{Fourier}

\begin{figure*}[!ht]
\centering
\includegraphics[width=0.85\textwidth]{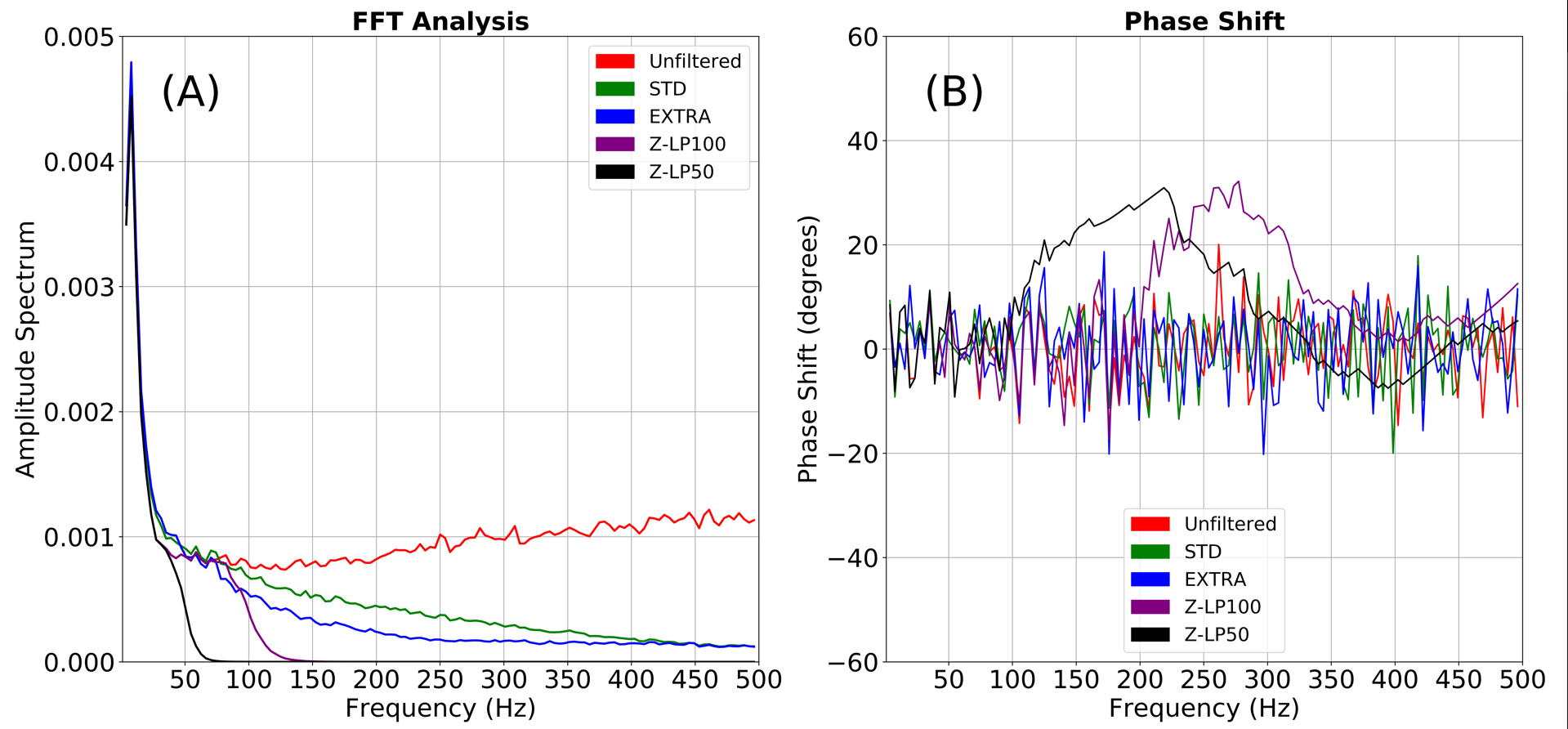}
\caption{Fourier analysis of the unfiltered signal and all of the filters evaluated in this report. For STD and EXTRA, data were filtered with our filter functions. (A) Amplitude spectra. (B) Phase shift spectra. Segments were chosen as described in \ref{segemntselection}}
\label{fig:3}
\end{figure*}

In Fig. \ref{fig:3}, we present the average amplitude spectra for all 5 signal types in plot (A).  Average phase-shift spectra are presented in plot (B).  

Amplitudes (plot (A)) for all signals are much higher in the very low frequencies (1-30 Hz).  All signals have much less amplitude above 100 Hz.  The amplitude of the unfiltered signal reaches a minimum around 150 Hz and then the amplitude actually increases as frequencies approach 500 Hz.  The STD filtered signal has a gradual decline in amplitude from about 50 Hz to 400 Hz, and then it flattens out to 500 Hz.  The EXTRA filtered signal reduces the amplitude of the signals to a lower level than the STD filter, especially in the range of 75-400 Hz.  The amplitude of the Z-LP100 filtered signal drops sharply at about 75 to 150 Hz, and remains essentially 0.000 above 150 Hz.  The amplitude of the Z-LP50 filtered signal drops sharply at about 30 Hz to 75  Hz and remains essentially 0.000 above 75 Hz.  

The phase-shift spectra generally indicates small phase-shift for most signals.  For the unfiltered, STD filtered and EXTRA filtered signals, we would expect that with more fixation segments from additional subjects, the phase shifts would approach 0.0.  Both the Z-LP100 and the Z-LP50 filters show systematic increases in phase shift. For the Z-LP50 filter, phase-shift begins to increase at about 100 Hz and decreases to a zero-level near 325 Hz. For the Z-LP100 filter, phase-shift begins to increase at about 200 Hz and then decreases to zero at about 400 Hz.  These phase-shift changes need to be interpreted with the knowledge that there was effectively no signal at these frequencies when the data are filtered with the L-ZP filters.  We think these increases in phase shift can practically be ignored.

\subsection{Signal frequency content analysis of eye-movements}
\subsubsection{Analysis of a Saccade with a PSO}
\begin{figure}[!ht]
\centering
  \includegraphics[width=0.7\textwidth]{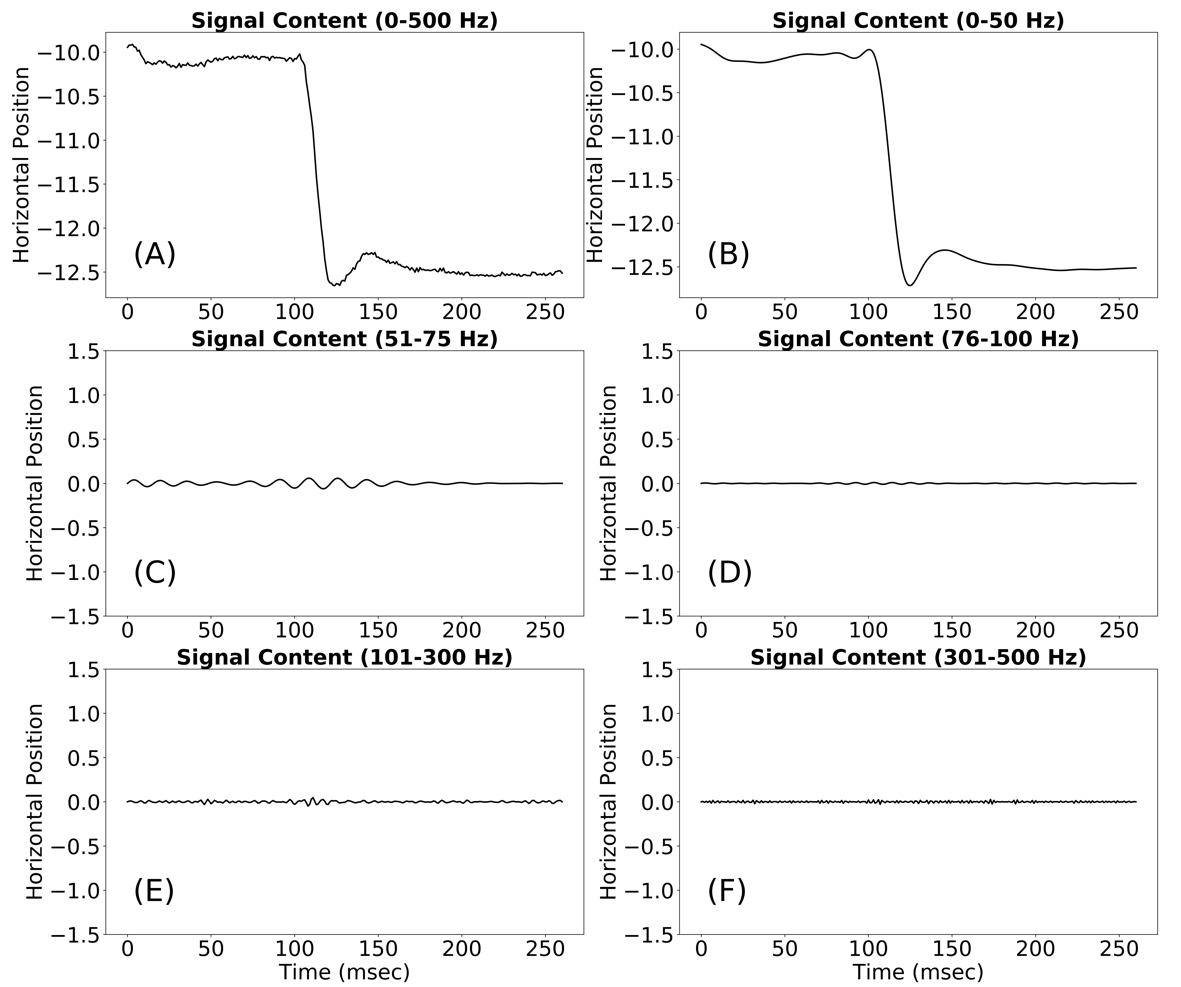}
\caption{Signal frequency content analysis of a saccade with a post-saccadic oscillation (PSO). (A) A sample unfiltered saccade with PSO signal. (B) The signal in (A) with only frequencies from 0 to 50 Hz. (C) The signal in (A) with only frequencies from 51 to 75 Hz. Note that all plots have the same amplitude range as the original saccade. (D) The signal in (A) with only frequencies from 76 to 100 Hz.(E) The signal in (A) with only frequencies from 101 to 300 Hz. (F) The signal in (A) with only frequencies from 301 to 500 Hz.}
\label{fig:4}
\end{figure}

\begin{figure}[!ht]
\centering
\includegraphics[width=0.7\textwidth]{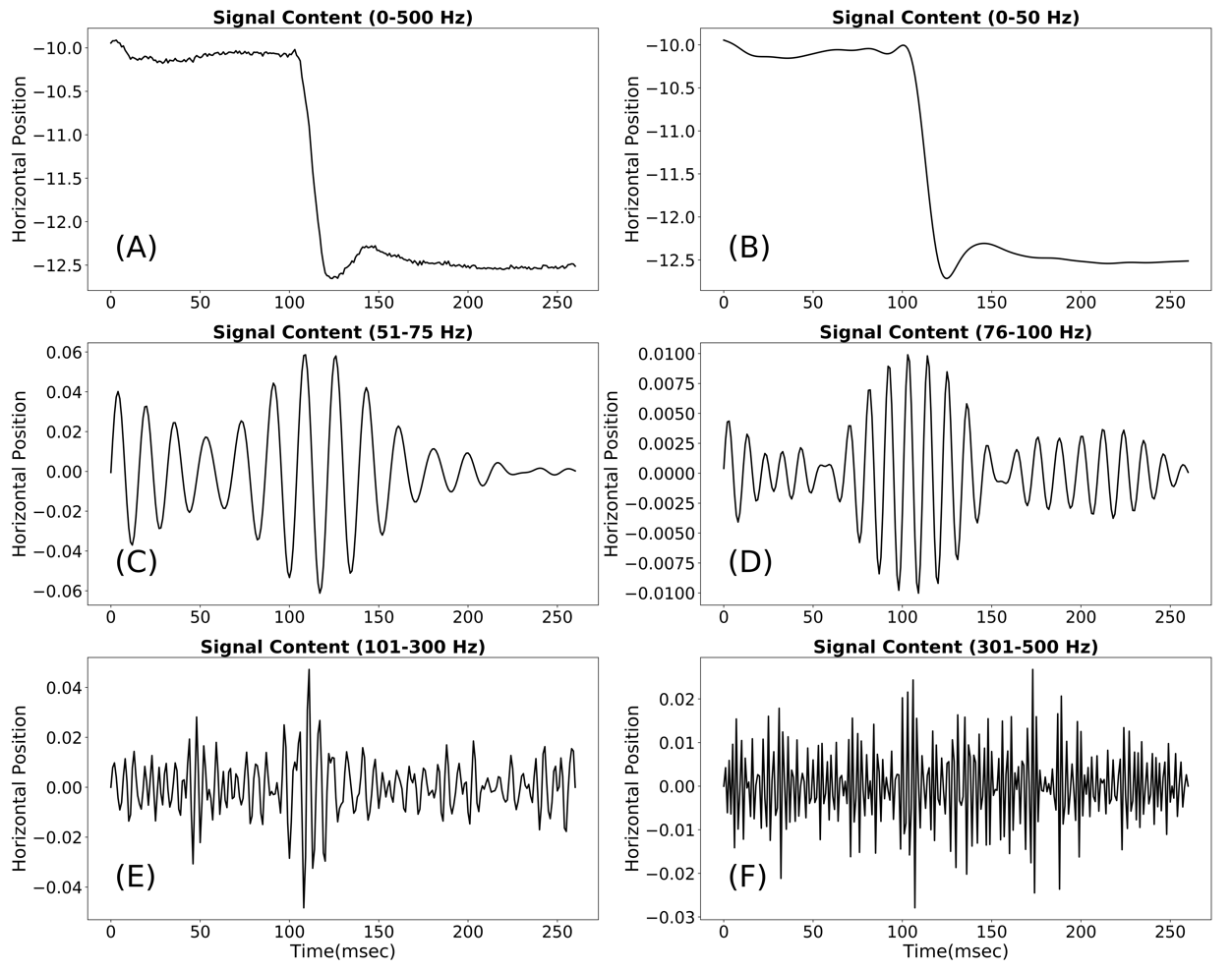}
\caption{Exactly the same analysis as in Fig. \ref{fig:4} except that each window is y-scaled individually according to the range of the data.}
\label{fig:5}
\end{figure}

In Fig. \ref{fig:4}, we present the signal frequency content analysis for a  saccade with a PSO. This saccade has an approximate amplitude of $2.7^o$.  In plot (A) we present the unfiltered signal trace for the saccade.  In plot (B) we present the signal containing frequencies from 0 to 50 Hz. You can see the saccade in plot (B) looks very similar to the saccade in plot (A) but with less noise. The saccade amplitude has not be altered. For this figure as well as Fig. \ref{fig:6} and  \ref{fig:8}, all a plots have the same position range as the unfiltered signal in plot (A).  Since you cannot see some of the signals on this scale very well, we also present figures with scales varying from plot to plot, optimized to view the signal content (Fig. \ref{fig:5}, Fig. \ref{fig:7} and Fig. \ref{fig:9}).

There was some minor contribution to signal amplitude in the 51-75 Hz range, but given how similar plot (B) is to plot(A), it would seem unlikely that we need this frequency range to accurately represent saccades.  There was essentially no contribution from any of the remaining frequency bands\footnote{For more examples of this and the other frequency content analyses see \url{https://digital.library.txstate.edu/handle/10877/16136}}.

\subsubsection{Analysis of catch-up saccade}
\begin{figure}[!ht]
\centering
\includegraphics[width=0.7\textwidth]{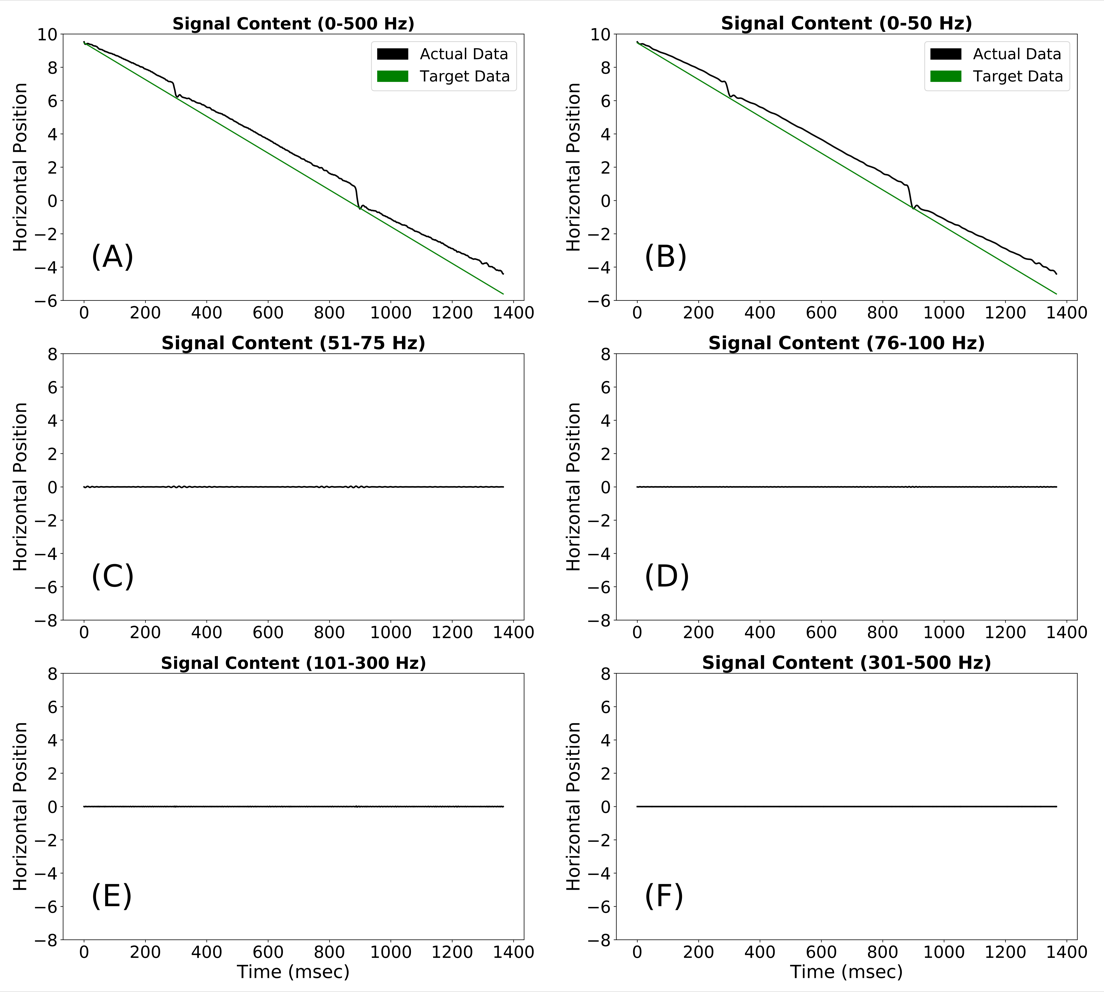}
\caption{Signal frequency content analysis of catch-up saccade. (A) A sample unfiltered catch-up saccade signal. (B) The signal in (A) with only frequencies from 0 to 50 Hz. (C) The signal in (A) with only frequencies from 51 to 75 Hz. Note that all plots have the same amplitude range as the original saccade. (D) The signal in (A) with only frequencies from 76 to 100 Hz.(E) The signal in (A) with only frequencies from 101 to 300 Hz. (F) The signal in (A) with only frequencies from 301 to 500 Hz.}
\label{fig:6}
\end{figure}

\begin{figure}[!ht]
\centering
  \includegraphics[width=0.7\textwidth]{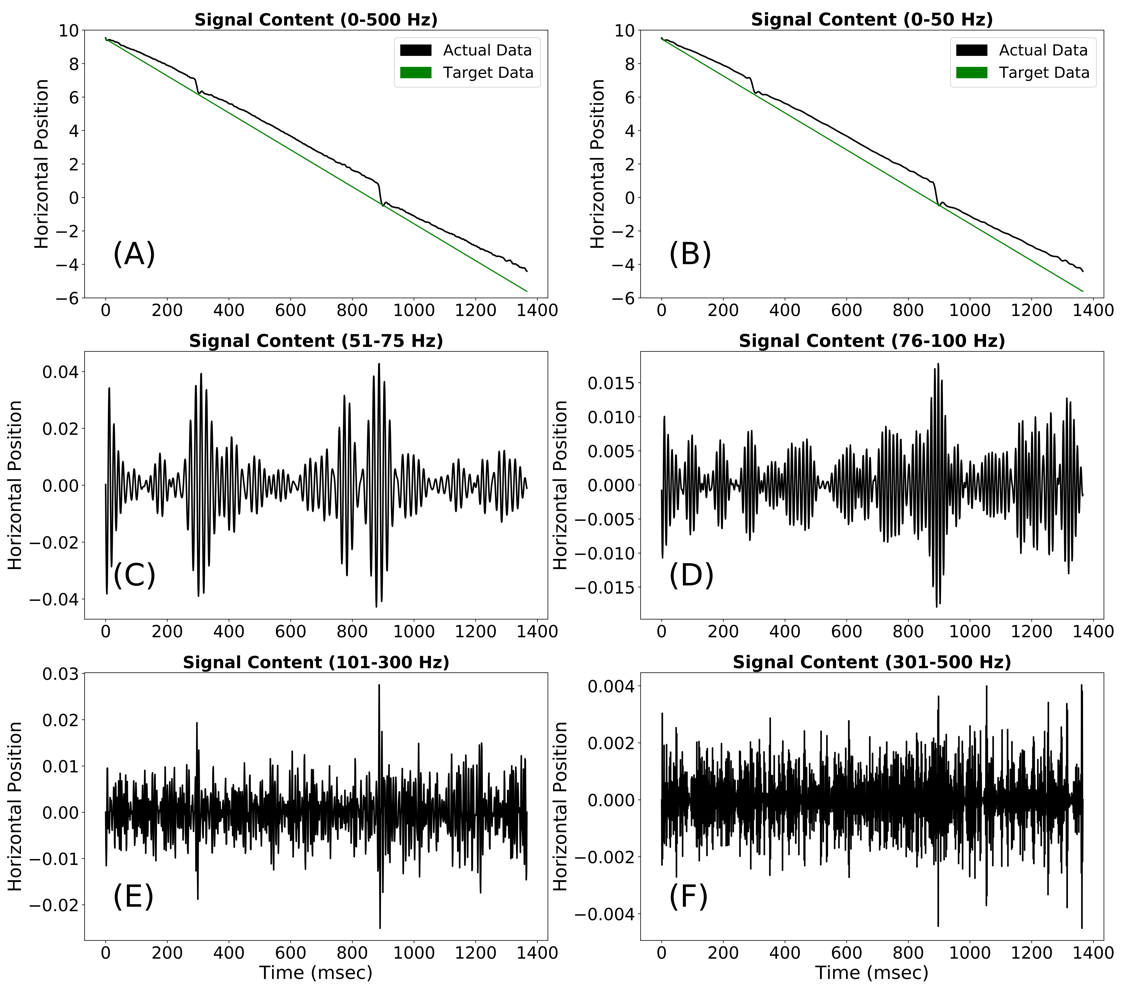}
\caption{Exactly the same analysis as in Fig. \ref{fig:6} except that each window is y-scaled individually according to the range of the data.}
\label{fig:7}
\end{figure}

In Fig. \ref{fig:6}, we present the signal frequency content analysis for 2 CUS, one about $0.92^o$ and one about $1.30^o$ in amplitude. The green line represents the target position. CUS occur when subjects have a gain of less than 1.0, and therefore the eye gradually falls behind the target.  To catch-up to the target, the subject makes CUS\footnote{For a detailed analysis of the relationship between smooth pursuit gain, CUS ampliutde and CUS rate see \citep{SmoothPUrsuitModel}}. Similar to the situation for the saccade with a PSO, the signal filtered from 0-50 Hz appears to contain all the data required to represent the CUSs with less noise.  The actual frequency content can best be judged in Fig. \ref{fig:7}.

\subsubsection{Analysis of a microsaccade}
\begin{figure}[!ht]
\centering
\includegraphics[width=0.7\textwidth]{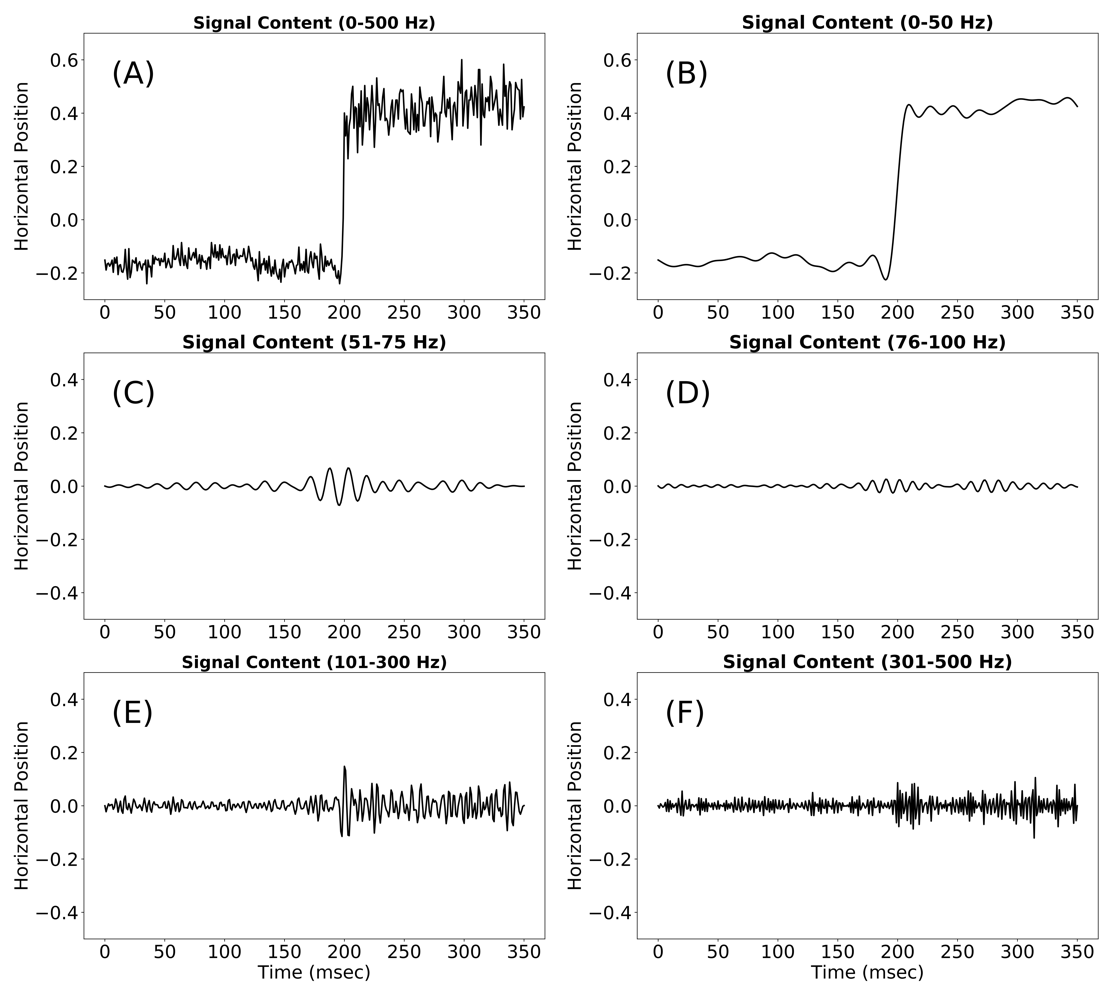}
\caption{Signal frequency content analysis of a microsaccade. (A) A sample unfiltered microsaccade. (B) The signal in (A) with only frequencies from 0 to 50 Hz. (C) The signal in (A) with only frequencies from 51 to 75 Hz. Note that all plots have the same amplitude range as the original saccade. (D) The signal in (A) with only frequencies from 76 to 100 Hz.(E) The signal in (A) with only frequencies from 101 to 300 Hz. (F) The signal in (A) with only frequencies from 301 to 500 Hz.}
\label{fig:8}
\end{figure}

\begin{figure}[!ht]
\centering
\includegraphics[width=0.7\textwidth]{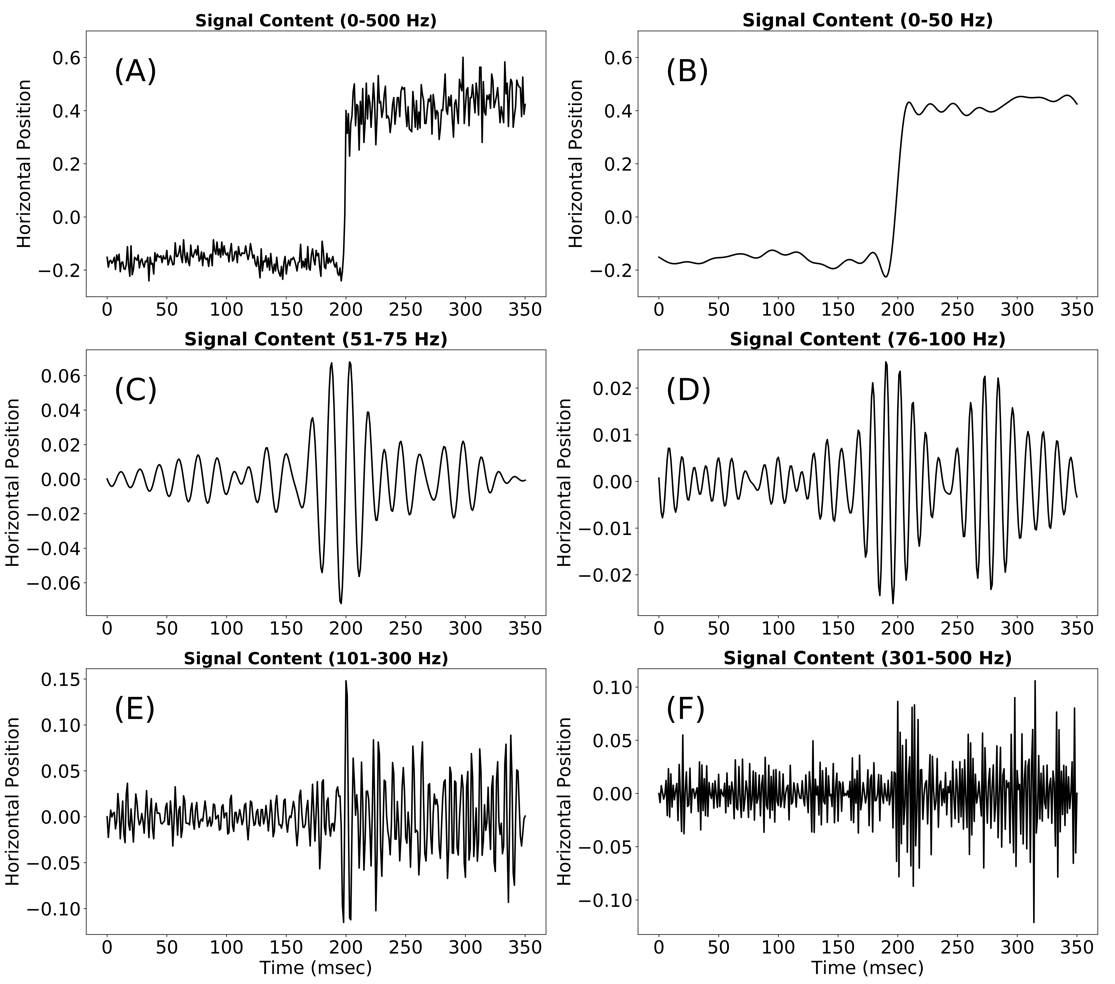}
\caption{Exactly the same analysis as in Fig. \ref{fig:8} except that each window was y-scaled individually according to the range of the data.}
\label{fig:9}
\end{figure}

In Fig. \ref{fig:8}, we present the signal frequency content analysis for a microsaccade. The amplitude of this microsaccade was approximately $0.6^o$. Although much of the signal was retained in plot (B), there was residual amplitude at higher frequencies. 

\subsection{Comparison of an unfiltered saccade eye movement signal with filtered versions of the same signal}\label{3in1}

\begin{figure*}[!ht]
\centering
\includegraphics[width=0.9\textwidth]{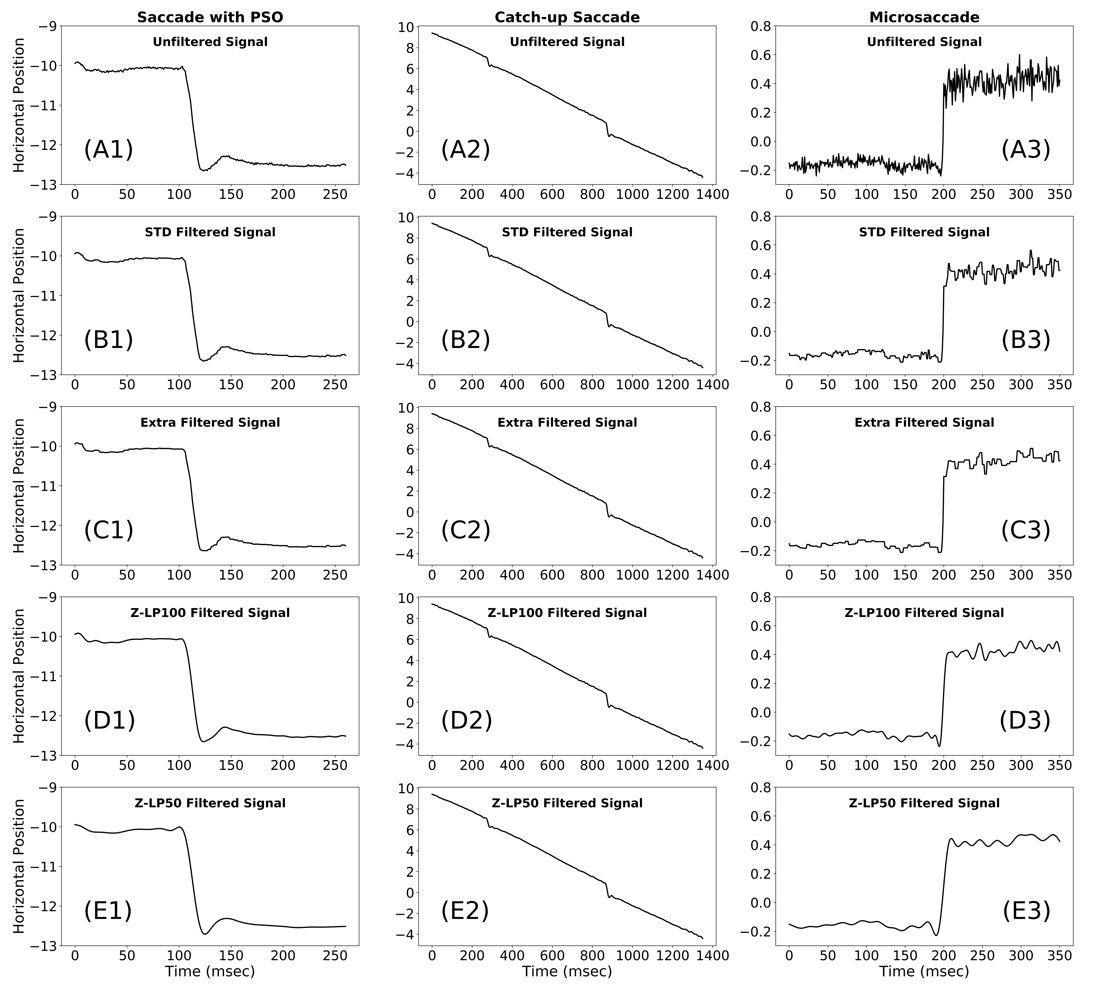}
\caption{Comparison of an unfiltered signal with filtered versions of the same eye-movement signal.
(A1) A sample unfiltered saccade with PSO signal.
(B1) Signal in (A1) has been filtered using our STD filter function.
(C1) Signal in (A1) has been filtered using EXTRA filter function
(which takes STD filtered data as input).
(D1) Z-LP100 filter has been applied to the signal in (A1).
(E1) Z-LP50 filter has been applied to the signal in (A1).
(A2) A sample unfiltered catch-up saccade signal.
(B2) Signal in (A2) has been filtered using our STD filter function.
(C2) Signal in (A2) has been filtered using EXTRA filter function
(which takes STD filtered data as input).
(D2) Z-LP100 filter has been applied to the signal in (A2). 
(E2) Z-LP50 filter has been applied to the signal in (A2).
(A3) A sample unfiltered microsaccade signal.
(B3) Signal in (A3) has been filtered using our STD filter function.
(C3) Signal in (A3) has been filtered using EXTRA filter function
(which takes STD filtered data as input).
(D3) Z-LP100 filter has been applied to the signal in (A3).
(E3) Z-LP50 filter has been applied to the signal in (A3).}
\label{fig:10}
\end{figure*}

In Fig. \ref{fig:10}, we present a comparison of unfiltered signal segments with the filtered versions of those signals.  We applied both the heuristic filters as provided by our EyeLink filter functions and our two Z-LP filters. In the low-noise figures in the left and middle column it is difficult to see much change.  The Z-LP50 filter does appear to round off some of the edges in the unfiltered signal.  All filters reduce high-frequency noise in this example microsaccade.

\subsection{Analysis of Filter Frequency Response}\label{fr}

\begin{figure*}[!ht]
\centering
\includegraphics[width=0.90\textwidth]{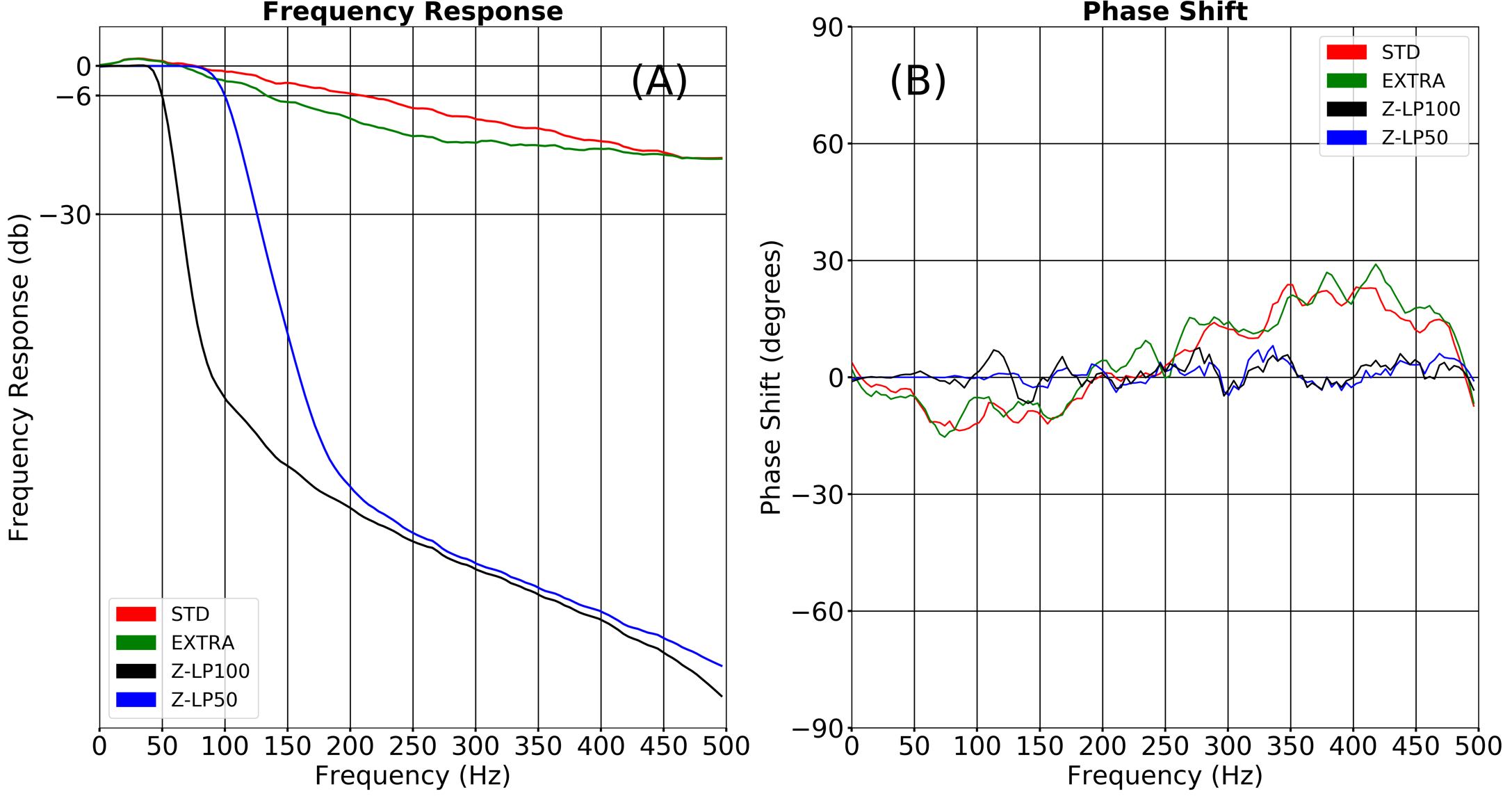}
\caption{Frequency response of all the filters (EyeLink filters, Digital filters) (A) Frequency response of STD, EXTRA, Z-LP100, and Z-LP50 filters (B) Phase shift of STD, EXTRA, Z-LP100, and Z-LP50 filters. }
\label{fig:11}
\end{figure*}

In Fig. \ref{fig:11} (A), we present the frequency response of STD, EXTRA, Z-LP100, and Z-LP50 filters. The Y-axis for plot (A) is in decibels (dB, reference = 1). The red line in (A) represents the frequency response of the STD filter. The green line represents the frequency response of the  EXTRA filter.  The black and blue line represents the frequency response of the Z-LP100 and Z-LP50 filters, respectively.  Obviously, the Z-LP filters remove much more data than the heuristic filters.  When a frequency response of a filter crosses the horizontal line at -6db, 25\% of the amplitude of original signal is retained.  When a frequency response moves below the horizontal line at -30db, only 0.1\% of the original signal remains.  Obviously the digital filters do a much better job at reducing higher frequencies without affecting the lower frequencies.  The heuristic filters remove high frequency signal much more slowly than the digital filters. The maximum amplitude of noise remaining at 500 Hz was -19 db (1.3 percent of signal remaining) for both of the heuristic filters.  For both of the Z-LP filters, at 500 Hz, the signal was effectively reduced to $0^o$ amplitude.

In Fig. \ref{fig:11} (B), the phase-shift caused by the filters remains close to 0.

\subsection{Effects of Filters on Saccade Peak Velocity}
\label{SlopeResult}
\begin{figure}[!ht]
\centering
\includegraphics[width=0.6\textwidth]{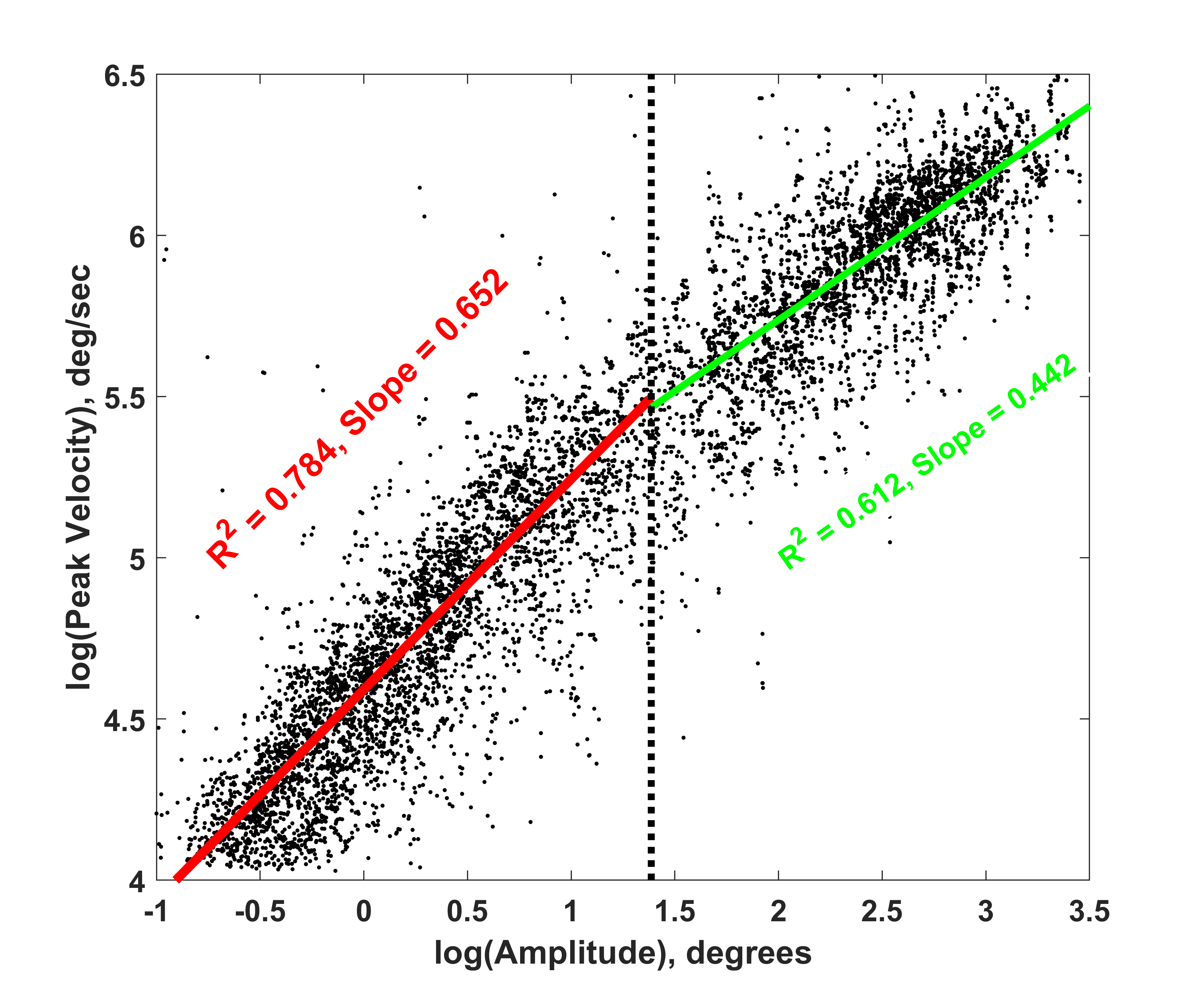}
\caption{Plot of log(peak velocity) against log(Saccade Amplitude). Data include saccades from 9 subjects across 5 filter conditions. These saccades are from our Random Saccade task and range from $0.36^o$ to $33.1^o$ for a total of 9,589 saccades. Two clusters of points, one less than or equal to $4^o$ (as logarithm, 1.39) of amplitude and one for larger saccades are apparent.  Fitted regression lines are shown in red and green.  The small saccade group slope was substantially greater than the slope for the large amplitude group.}
\label{fig:12}
\end{figure}

In Fig. \ref{fig:12}, we present a log(peak velocity) versus log-(saccade amplitude) plot for every saccade from our Random Saccade task. These saccades range in size from $0.38^o$ to $33.1^o$.  Note the presence of two clusters of data, one with saccade amplitude less than or equal to $4^o$ (as logarithm, 1.39) and one larger.  The slope for the small cluster was substantially greater than the for the large cluster.

In Table \ref{tab:slopes}, we present our analysis of the slopes of small ($<=4^o$) saccades and large saccades from our random saccade task.  The table contains the mean (SD) slopes for small and large saccades.  For the small saccades, the lowest slope was for the EXTRA filtered data and the highest slope was for the Z-LP50 filter.  There was a statistically significant fixed effect for filter (F=3.16, df=40, p=0.024).  Post-hoc analysis revealed that the slope for the Z-LP50 data was significantly greater than all the other filter conditions. For the large saccades, the filter effect was not statistically significant (F=0.036, df=40, p = 0.997).  Leaving the Z-LP50 filter aside, no filter condition was statistically significantly different from unfiltered data.

\begin{table}[!ht]
    \centering
    \caption{Slopes for Small and Large Saccades}
    \label{tab:slopes}
    \begin{tabular}{|c|c|c|c|c|}
    \hline
        Filter & Mean & SD & Mean & SD \\ 
               & Small & Small & Large & Large \\ \hline
        No Filter & 0.673 & 0.089 & 0.438 & 0.117 \\ \hline
        STD & 0.670 & 0.086 & 0.445 & 0.115 \\ \hline
        EXTRA & 0.638 & 0.084 & 0.445 & 0.111 \\ \hline
        Z-LP100 & 0.686 & 0.101 & 0.439 & 0.122 \\ \hline
        Z-LP50 & 0.780 & 0.092 & 0.456 & 0.108 \\ \hline
    \end{tabular}
\end{table}

In Table \ref{tab:HOR}, we present the analysis of the large saccades ($22^o$ to $35.4^o$) from out horizontal saccade task.  The table has the mean (SD) amplitude and the mean (SD) peak velocity for each filter group.  The filter types did not differ for saccade amplitude (F=0.047, df=5340, p=0.997).  The Table shows that the Z-LP50 filter group has the lowest peak velocity and the unfiltered data has the largest peak velocity.  The filter type effect was statistically significant (F=8.66, df=40, $p < 0.001$ ).  Post hoc analysis revealed that the peak velocity for the Z-LP50 filter was statistically significantly lower than the unfiltered data, the STD filtered data and the EXTRA filtered data.  Leaving the Z-LP50 filter aside, no filter condition was statistically significantly different from unfiltered data.  

\begin{table}[!ht]
    \centering
    \caption{Saccade Amplitudes and Peak Velocities for Very Large Saccades}
    \label{tab:HOR}
    \begin{tabular}{|c|c|c|c|c|c|}
    \hline
        FiltGrp & Count & Mean & SD & Mean & SD \\ 
                &       & Ampl & Ampl & PkVel & PkVel \\ \hline
        No Filt & 1059 & 28.26 & 2.09 & 531.15 & 77.01 \\ \hline
        STD & 1069 & 28.26 & 2.10 & 530.82 & 77.50 \\ \hline
        EXTRA & 1084 & 28.24 & 2.09 & 530.46 & 77.50 \\ \hline
        Z-LP100 & 1084 & 28.27 & 2.10 & 523.21 & 78.14 \\ \hline
        Z-LP50 & 1049 & 28.28 & 2.11 & 515.08 & 77.05 \\ \hline
    \end{tabular}
\end{table}

\section{Discussion}
\label{discuss}

For those who use the their video-oculography-based eye-trackers as a data collection device only, we have supplied two tools that will be very helpful.  First, we have supplied software functions which perform the heuristic filters \citep{stampe} that several eye-trackers use. 
This approach is more versatile as it provides the user with unfiltered, STD filtered and EXTRA filtered data simultaneously, from the exact same recording.  In this way, investigators can decide which type of filtering is best for a specific application.  For studies which compare the heuristic filter levels (e.g., \citep{ArtificalEye}) these functions will eliminate the need (and associated interpretation issues) of collecting multiple recordings from the same subject at different filter levels.  Second, we have shown that the frequency response of the STD and EXTRA filters are inferior to the frequency response of our proposed zero-phase digital low pass filter (Z-LP100).  The digital filter more completely suppresses the high frequencies that are not needed. 

Of course, future investigators can adjust the cutoff frequency (100 Hz) or choose a different base filter (Butterworth) for their own low pass filters.  We believe that frequencies above 100 Hz are not needed for most eye-movement research performed with video-based systems, but others may differ and choose a higher (e.g., 125 Hz) cutoff frequency.

However, according to \url{https://community.sw.siemens.com/s/article/digital-signal-processing-sampling-rates-bandwidth-spectral-lines-and-more}:

\begin{quote}
``To get close to the correct peak amplitude in the time domain, it is important to sample at least 10 times faster than the highest frequency of interest. For a 100 Hertz sine wave, the minimum sampling rate would be 1000 samples per second. In practice, sampling even higher than 10x helps measure the amplitude correctly in the time domain.''
\end{quote}

Given that our data were collected at 1000 Hz, and we want to preserve up to 100 Hz data, we need a sampling frequency of 1000 Hz.  To preserve 125 Hz data, as sampling rate of 1250 Hz would be needed, at a minimum.

Stampe \citep{stampe} rejected digital low-pass filters in favor of heuristic filters, because he thought that video-based eye-tracking systems had low sample rates.  This may have been true in 1993, but it is not true today.  He predicted, on the basis of no evidence, that digital low-pass filters would negatively affect saccade detection but we see no data suggesting that this is so.  In fact, better filtering should make the detection of all eye-movements easier, but this is an empirical question. We are very comfortable recommending that, in future, these heuristic filters be replaced by digital low-pass filters.

Wierts et al \citep{50Hz} found that, for saccades up to $5^o$, peak velocities are preserved with a sampling rate of 50Hz.  In the present study (Section \ref{SlopeResult}) we tested the effect of heuristic and digital low-pass filters on saccade peak velocities.
The Z-LP50 filter (which has a cut-off frequency of 50 Hz) had some effects on saccade peak velocity.  Regardless of saccade size, we found no evidence that any other filter (STD, EXTRA, Z-LP100) altered saccade peak velocities, as compared to the unfiltered condition for saccades up to $35.4^o$ degrees.  

As we noted in introduction, fixation microtremor is not typically studied with an EyeLink 1000 (or any video-based) system.  Typically studies which evaluate microtremor find that these movements have frequency content between 50 Hz and 95 Hz.  It would seem that a cutoff in the range of 100-150 Hz would be required to get an accurate evaluation of this phenomenon.  

We have not evaluated the effect of filtering on studies of nystagmus or the vestibular optokinetic reflex or other eye-movements.  Therefore, we currently have no recommendations regarding these movements.  We did evaluate recordings of CUS during smooth pursuit and found that the frequencies needed for typical saccades apply also to CUS.

In the future, it might be interesting to perform the same type of study using other popular eye-tracking systems. Perhaps our analysis would yield different results for different systems.  For the present, our results apply to EyeLink eye-trackers only.

\section*{Acknowledgments}
The authors declare no conflict of interest. The funders had no role in the design of the study; in the collection, analyses, or interpretation of data; in the writing of the manuscript, or in the decision to publish the results.

\section*{Open Practices Statement}
The data and code for the present study is available at \url{https://digital.library.txstate.edu/handle/10877/16136}.
Download Supplementary\_Materials.zip. This study was not pre-registered.

\section*{Conflict of interest}
The authors declare no conflict of interest. The funders had no role in the design of the study; in the collection, analyses, or interpretation of data; in the writing of the manuscript, or in the decision to publish the results.

\bibliographystyle{unsrtnat}
\bibliography{Fourier.bib}  

\end{document}